\documentclass{article}
\usepackage{ijcai22}

\usepackage{times}
\usepackage{soul}
\usepackage{url}
\usepackage[hidelinks]{hyperref}
\usepackage[utf8]{inputenc}
\usepackage[small]{caption}
\usepackage{graphicx}
\usepackage{amsmath}
\usepackage{amsthm}
\usepackage{booktabs}
\usepackage{algorithm}
\usepackage{algorithmic}
\usepackage[np]{numprint}
\npthousandsep{,}

\usepackage[inline]{enumitem}

\newcommand{\fsf}[1]{{\small\textsf{#1}}}

\newcommand{\fsub}{\textsubscript}

\DeclareMathAlphabet{\mathsl}{OT1}{ptm}{m}{sl}

\usepackage{colortbl}

\usepackage{siunitx}
\usepackage{multirow}
\graphicspath{{./images/}}

\usepackage{xcolor}

\definecolor{tabblue}{RGB}{31,119,180}
\definecolor{taborange}{RGB}{255,127,14}

\usepackage{subcaption}

\usepackage{pgfplots}
\pgfplotsset{compat=newest}
\usetikzlibrary{pgfplots.statistics}
\usepgfplotslibrary{groupplots}

\title{Socially Intelligent Genetic Agents for the Emergence of Explicit Norms}
\author{Rishabh Agrawal$^{1}$
\and 
Nirav Ajmeri$^{2}$
\And 
Munindar~P. Singh$^{1}$\\
\affiliations
$^{1}$North Carolina State University\\
$^{2}$University of Bristol\\
\emails
{ragrawa3@ncsu.edu, nirav.ajmeri@bristol.ac.uk, mpsingh@ncsu.edu}
}

\newcommand{\citep}{\cite}
\newcommand{\citet}[1]{\citeauthor{#1} [\citeyear{#1}]}
\newcommand{\citepos}[1]{\citeauthor{#1}'s [\citeyear{#1}]}

\begin{document}

\maketitle

\begin{abstract}
Norms help regulate a society. Norms may be explicit (represented in structured form) or implicit. 
We address the emergence of explicit norms by developing agents who provide and reason about explanations for norm violations in deciding sanctions and identifying alternative norms. These agents use a genetic algorithm to produce norms and reinforcement learning to learn the values of these norms.
We find that applying explanations leads to norms that provide better cohesion and goal satisfaction for the agents. Our results are stable for societies with differing attitudes of generosity.
\end{abstract}

\section{Introduction}
\label{sec:intro}

Norms encourage coordination and prosocial interactions in a society \citep{Morris+19:norm-emergence}. For example, ignoring a phone call in a meeting is a social norm that helps avoid disruption.  
Importantly, norms may conflict with one another \citep{kollingbaum2007managing,Santos-JAAMAS17-NormConflictSurvey}: an agent must decide which norms to follow and which to violate. For example, \emph{picking up an urgent call} and \emph{ignoring a call during a meeting} are both norms. But if you receive an urgent call during a meeting, one or the other norm must be violated.

Norms can be \emph{implicit} (encoded in common behaviors \citep{Morris+19:norm-emergence}) or \emph{explicit} (explicitly maintained and reasoned about like laws \citep{Von-Wright-63:Norm} and regulations \citep{Jones+Sergot-93,TOSEM-20:Desen}). 
Where do norms come from? 
Prior work considers (decentralized) \emph{emergence} for implicit norms  and (central) \emph{synthesis} for explicit ones \citep{Morales-AAMAS14-Minimality,Morales2018Offline}.

\emph{Norm emergence} is the decentralized evolution of norms \citep{Morris+19:norm-emergence}, driven by signaling between agents.
A norm violation may result in a negative \emph{sanction} \citep{Nardin+16:Sanctioning}, i.e., a scowl on the face of a meeting attendee. 
If the scowl is outweighed by the benefits of picking up an urgent call, the norm \emph{ignore a call during a meeting, unless it is an urgent call} may emerge. Alternatively, if the sanction is severe, the norm \emph{pick up an urgent call, unless you are in a meeting} may emerge. For us, a norm violation is not necessarily considered a negative act, but rather a decision by an autonomous agent to choose between norms. Thus, norm violations drive norm change \citep{castelfranchi2013cognitive}.

We focus on the emergence of \emph{explicit} norms.
\citet{Ajmeri-IJCAI18-Poros} show that sharing contextual information in case of norm violations facilitates norm emergence for implicit norms and results in increased goal satisfaction and cohesion (i.e., perception of norm compliance) by helping agents understand the context (i.e., attributes that define the circumstances) in which the norm was violated and thus learn contextual boundaries.
By sharing \emph{explanations} in cases of norm violations, agents can continually refine existing norms. Examples include photo sharing \citep{mosca2021elvira} and user acceptance in general \citep{ye1995impact}. Explanations underlie accountability, and refining norms yields innovation in a sociotechnical system \citep{WWW-16:IOSE}.

Our research objective is to understand how creating and sharing \emph{explanations} for norm violations facilitates the emergence of norms.
We develop agents who produce and reason with explanations. Explicit norms are conducive to explanations of when they are satisfied or violated. 
We propose a method for the emergence of explicit norms by developing genetic agents that use rule learning \citep{liu2016rule}.
Based on the foregoing, we identify these research questions:

\begin{description}
\item[RQ\fsub{G} (Goal)]
\emph{Do societies of agents who provide and evaluate explanations for norm violations achieve higher goal satisfaction for their members than other societies?}

\item[RQ\fsub{C} (Cohesion)] \emph{Does providing and evaluating explanations lead to norms that improve social cohesion?}

\end{description}

We make these contributions.
First, we develop a socially intelligent genetic agent (SIGA) in two variants: A plain SIGA (or NSIGA) represents norms explicitly; an XSIGA in addition explains its actions and incorporates others' explanations.
Second, using SIGAs, we answer both RQs positively: (1) Sharing explanations to explain actions lead to improved goal satisfaction; and (2) sharing explanations to explain actions lead to emergence of norms with increased cohesion.

\section{Running Example and Solution Idea}

In our model, each agent has one \emph{primary} stakeholder, whom the agent represents, and potentially many \emph{other} stakeholders who are affected by the agent's actions \citep{AAMAS-20:Elessar}. 

We adopt a phone ringer application to explain our method.
An agent is responsible for ringing the phone of its primary stakeholder (the callee) when a call is received or to keep it silent. The other stakeholders are the caller and the people in the vicinity of the callee who may be disturbed by the ringing phone. Actions available to the agent are \emph{ring} or \emph{ignore}. The agent decides to ring the phone or keep it silent based on the norms it follows. Agents of other stakeholders may sanction the agent if they don't agree with its action. 

Suppose Alice's agent follows three norms: \emph{always ring an urgent call}, \emph{always ring a call from a family member}, and \emph{ignore calls during a meeting}. 
Suppose Bob's agent follows the norms: \emph{always ring an urgent call} and \emph{ignore calls during a meeting}. 
Suppose Alice gets an urgent call from a family member during a meeting. Her agent assigns a value (expected reward) to each action in terms of the norms that support it. 
If the combined value of \emph{always ring an urgent call} and \emph{always ring a call from a family member} is more than the norm \emph{ignore calls during a meeting}, it rings the phone. 

If Alice's agent does not explain its action, Bob's agent would assume that the only applicable norm, \emph{ignore calls during a meeting}, was violated, and sanction Alice. 
If Alice presents the supporting norms, \emph{always ring an urgent call} and \emph{always ring a call from a family member}, they are evaluated by Bob's agent. 
Suppose Charlie's agent does not follow \emph{always ring a call from a family member} but follows \emph{always ring an urgent call}. 
If Charlie's agent assigns a higher value to \emph{always ring an urgent call} over \emph{ignore calls during a meeting}, it would not issue a sanction. Else, it would reject the explanation and sanction Alice, leading her agent to adjust its values of the three norms.

\section{Method: Realizing a SIGA}
\label{sec:proposed_method}

A SIGA's actions are governed by explicit norms. 
Norms in our conception are commitments \citep{Singh-2013-Norms}.
A commitment is written \fsf{Commitment(subject, object, \emph{antecedent}, \emph{consequent})}. The antecedent determines when it goes in force; the consequent determines when it completes (is satisfied or violated); the subject is the agent who commits, and the object is the agent to whom it is committed.
We can express \emph{(Callee) commits to always ringing a call from a family member (Caller)} as:
\fsf{Commitment}$(\mbox{Callee}, \mbox{Caller}, \mbox{callerRel = \emph{family}}, \mbox{action=\emph{ring}})$.
Commitments are well suited for rule-based learning as rules can suggest actions that can map to a commitment's consequent.

\subsection{Norm Learning and Discovery}
\label{sec:norm-learning}

A rule-based approach to implementing norms supports flexibility and ease of implementation. Each agent stores norms as rules that it learns, evaluates, and evolves. (Below, we write $\top$ for true and $\bot$ for false.)
Here, the antecedents are conjunctions of key-value pairs, e.g., callerRel = family, urgent = $\top$. The consequents are actions to be taken to satisfy the norm. Thus, a norm maps to a rule of the form:
%
    $\mbox{IF } \mbox{\emph{antecedent}} \mbox{ THEN } \mbox{\emph{consequent}}$.

We adapt \emph{eXtended Learning Classifiers} (XCS), a rule learning algorithm based on reinforcement learning \citep{butz2000algorithmic,urbanowicz2017introduction}, which evolves rules by learning from rewards obtained from the environment in response to actions. This algorithm enables agents to discover norms and learn their values.
Specifically, we 
\begin{enumerate*}[label=(\arabic*)]
\item aggregate the sanctions and payoffs received by an agent into rewards;
\item map norms to IF-THEN rules to be manipulated by XCS; and
\item define crossover and mutation operations for norms to enable norm discovery.
\end{enumerate*}

XCS operates in two modes, \emph{exploration} (take random actions to find new norms) and \emph{exploitation} (apply learned rules and their weights to choose an action). We adopt the $\epsilon$-greedy technique, choosing exploration with probability $\epsilon$.

Each rule has associated parameters of \emph{fitness} (worth of the rule in making accurate predictions), \emph{reward prediction}, \emph{prediction error} (of the reward), and \emph{numerosity} (how many ``copies'' of the rule exist, indicating robustness against accidental deletion \citep{urbanowicz2017introduction}).

\subsubsection{Create Match Set}
The \emph{match set} is the set of rules that are activated in a given context. This means that the antecedents of the corresponding norms are true in the given context. 

\subsubsection{Cover Context}
Covering ensures that rules with sufficiently many actions are available for a given context. Covering maintains diversity by adding new rules to the agent's ruleset, to avoid overfitting to the initial conditions. 
The new rules are generated by randomly selecting some subset of the context to be the antecedent of the rule. This ensures that the antecedent would be true in the given context. The consequent is randomly selected from available actions.
Covering is essential at the beginning since each agent has an empty ruleset, meaning its match set is empty.

\subsubsection{Select Action}
The expected value of each action aggregates the fitness-weighted reward predictions of rules supporting that action---to predict the expected reward of following each norm. Pick the available action with the highest expected value.

\subsubsection{Create Action Set}
Matching rules that support a chosen action form the \emph{action set}.
We create this set by identifying which rules had suggested the selected action.

\subsubsection{Update Rule Parameters}
These updates follow \citet{urbanowicz2017introduction}.
Upon receiving a reward, the agent revises its value estimation of the rules by updating its rule parameters. Only rules in the action set are updated because we don't know the influence of selecting any other action.
Equation~\ref{eq:p_update} updates the predicted reward. Here, $p$ is the reward prediction, $r$ is the reward received, and $\beta$ is the rate of learning hyperparameter.
\begin{equation}
\label{eq:p_update}
    p \gets p + \beta (r - p)
\end{equation}

Equation~\ref{eq:varepsilon_update} updates the prediction error, $\varepsilon$. 
Equation~\ref{eq:accuracy} estimates accuracy $\kappa$ where $\varepsilon_{0}$ is the error threshold below which we assume a rule to be accurate. Here, $v$ controls the relationship between error and accuracy to increase the difference in fitness levels between two rules that are close in prediction error---to prefer a less error-prone rule during discovery. And, $\alpha$ is the scaling factor used to raise the least error-prone non-accurate classifier to be close to an accurate classifier.
\begin{equation}
\label{eq:varepsilon_update}
    \varepsilon \gets \varepsilon + \beta (|r - p| - \varepsilon)
\end{equation}
\begin{equation}
\label{eq:accuracy}
\kappa = 
    \begin{cases}
    1 & \mbox{if  } \varepsilon < \varepsilon_{0}\\
    
    \alpha(\frac{\varepsilon}{\varepsilon_{0}})^{-v} & \mbox{otherwise}
    
    \end{cases}
\end{equation}

We normalize accuracy to $\kappa'$ using Equation~\ref{eq:accuracy_normalization}, and update fitness $F$ using the \citet{widrow1960adaptive} update:
\begin{equation}
\label{eq:accuracy_normalization}
    \kappa' = \frac{\kappa}{\sum_{cl \in |A|} \kappa_{cl}}
\end{equation}
\begin{equation}
    F \gets F + \beta(\kappa' - F)
\end{equation}

\subsubsection{Subsume Action Set Rules}
Subsumption means replacing a rule with a more general rule that yields a smaller \emph{prediction error}---as defined in Section~\ref{sec:norm-learning}, the error in the reward predicted to be received when following the rule.
This process provides generalization pressure to the algorithm.
If the first rule below (R1) is more error-prone than the second (R2), it may be replaced by the second, thus increasing R2's numerosity (see Section~\ref{sec:norm-learning}).
\begin{align*}
\mbox{IF urgent=$\top$ }\land \mbox{callerRel = friend  } & \mbox{THEN ring} \tag{R1}\\
\mbox{IF urgent=$\top$  } & \mbox{THEN ring} \tag{R2}
\end{align*}

\subsubsection{Discover Rules}
We apply a genetic algorithm (GA) to generate new rules from current rules of high fitness. GA selects parents using \emph{tournament selection} \citep{urbanowicz2017introduction}.
We randomly select 30\% of the rules (with replacement, so they can be the same rule) in the action set to compete. The fittest two among them become parents. We generate two children using \emph{crossover} and \emph{mutation} operations, which we define for norms.
The children are added to the population of rules maintained by the agent.
To allow the rules to stabilize, we breed them only when the average experience (number of times a rule has been selected) of the rules in the action set is above a threshold.
We use single-point crossover, in which the values are randomly swapped for each contextual property present in either parent's antecedent.

In XCS, mutation is used to create a more general or more specific rule by randomly flipping bits in the parent encoding. For norms, we randomly add (using $\land$) or remove key-value pairs in the antecedent of the norm being mutated. 
For example, if the antecedent of a norm is \{callerRel = friend $\land$ urgent = $\top$\}, we may mutate it to \{urgent = $\top$\} by removing a pair or to \{callerRel = friend $\land$ urgent = $\top$ $\land$ calleeLoc = home\}, if the location in the current context is home.

\subsubsection{Subsume Child Rules}
Upon creation, a child rule is subsumed into its parent if the parent is more general and its error is less than a threshold. 

\subsubsection{Delete Rules}
A hyperparameter defines the maximum number of rules a SIGA can keep. We apply \citepos{kovacs1999deletion} deletion Scheme~3, which prefers to delete unfit rules.

\subsection{Norms as Explanations}

An explanation is comprised of norms that support the action taken, i.e., the action set identified above.
An agent evaluating the explanation identifies norms it follows in the explanation plus the norms that have been violated. It adds the associated rules to the match set and then follows the same procedure as action selection by performing fitness-weighted aggregation of reward prediction. If the resulting action matches the observed action, it applies no sanctions.

\section{Simulation Scenario: RINGER}
\label{sec:sim}

This scenario is based on our running example and is implemented using MASON \citep{Luke-2005-Mason}. 
Our simulation consists of a population of agents. There are five shared locations where agents can interact: homes (H), parties (P), meetings (M), a library (L), and an emergency room (ER). Some of these locations (H, P, and M) have an associated relationship circle. Each home has a family circle, each party has a friend circle, and each meeting has a colleague circle. People of the same circle share that relationship. An agent stays at one location for a random number of steps chosen from a Gaussian distribution with a mean of 60 steps and a standard deviation of 30, with the number of steps restricted to the range $[30, 90]$. Then it moves to another location. An agent is more likely to enter a location that is associated with its own circles (75\% probability) than a location with which it has no association (25\% probability). For example, an agent is more likely to enter its own home than a stranger's home.

At each timestep, an agent calls another agent with a probability chosen from a Gaussian distribution with a mean of 5\% and a standard deviation of 1\%. There is a 25\% probability each of calling a family member, a colleague, a friend, or a stranger.

Each agents has goals based on its role (Table~\ref{tab:agent_goals}). The degree to which an agent is affected by the promotion or demotion of these goals determines the payoff that it receives. The payoff may differ based on location and relationship.

\begin{table}[t]
\centering
     \begin{tabular}{@{}l@{~~} p{5.5cm} @{~~} l@{}} 
     \toprule
     \textbf{Role} & \textbf{Goal} & \textbf{Action}\\
     \midrule
     Caller & Reach callee either urgently or casually & Answer\\\midrule 
     
     \multirow{3}{*}{Callee} & To be reachable & Answer\\ 
     & To not be disturbed & Ignore\\ 
     & To not disturb others & Ignore\\\midrule 
     
     Neighbor & To not be disturbed & Ignore\\
     \bottomrule
    \end{tabular}
\caption{Agent goals.}
\label{tab:agent_goals}
\end{table}

Tables~\ref{tab:callee_payoff}, \ref{tab:caller_payoff}, and~\ref{tab:neighbor_payoff_fixed} present the callee, caller, and neighbor payoffs, respectively.
Neighbor payoffs and sanctions may be affected by the explanation provided, depending on whether the explanation was accepted or not. Table~\ref{tab:neighbor_payoff_explanation} summarizes the expected payoff of each situation in this case.

\begin{table}[t]
\centering
     \begin{tabular}{@{}p{3.0cm} l S S@{}} 
     \toprule
     \textbf{Caller Relationship} & \textbf{Callee Action} & \textbf{Casual} & \textbf{Urgent}\\
     \midrule
     Family, Friend, or & Answer & 0.50 & 1.00\\ 
     Colleague & Ignore & 0.00 & -0.50\\ 
     \midrule
     \multirow{2}{*}{Stranger} & Answer & -1.50 & 0.50\\ 
     & Ignore & 1.50 & -0.25\\ 
     \bottomrule
    \end{tabular}
\caption{Callee payoff based on urgency and relationship.}
\label{tab:callee_payoff}    
\end{table}

\begin{table}[t]
\centering
     \begin{tabular}{l S S} 
     \toprule
     \textbf{Callee Action} & \textbf{Casual} & \textbf{Urgent}\\
     \midrule
     Answer & 0.50 & 1.00\\ 
     Ignore & -0.50 & -1.00\\ 
     \bottomrule
    \end{tabular}
\caption{Caller payoff based on urgency of the call.}
\label{tab:caller_payoff}
\end{table}

\begin{table}[t]
\centering
     \begin{tabular}{@{}l@{~} S @{~~} S @{~~} S @{~~} S @{~~} S@{}} 
     \toprule
     \textbf{Callee's Action} & \textbf{ER} & \textbf{H} & \textbf{L} & \textbf{M} & \textbf{P}\\
     \midrule
     Answer & 1.00 & 0.67 & -1.00 & -1.00 & -0.33\\ 
     Ignore & -1.00 & -0.33 & 1.00 & 1.00 & 0.67\\ 
     \bottomrule
    \end{tabular}
\caption{Neighbor payoff based on location of the call.}
\label{tab:neighbor_payoff_fixed}
\end{table}

\begin{table}[t]
\centering
     \begin{tabular}{@{}p{1.0cm} @{~~} p{1.2cm} @{~~} S @{~~} S @{~~} S @{~~} S @{~~} S@{}}
     \toprule
     \textbf{Callee Action} & \textbf{Neighbor Expects} & \textbf{ER} & \textbf{H} & \textbf{L} & \textbf{M} & \textbf{P}\\
     \midrule
     Answer & Answer & 1.00 & 0.67 & 1.00 & 1.00 & 0.67\\ 
     Answer & Ignore & -1.00 & -0.33 & -1.00 & -1.00 & -0.33\\ 
     Ignore & Answer & -1.00 & -0.33 & -1.00 & -1.00 & -0.33\\ 
     Ignore & Ignore & 1.00 & 0.67 & 1.00 & 1.00 & 0.67\\ 
     \bottomrule
    \end{tabular}
\caption{Neighbor payoff based on evaluated explanations.}
\label{tab:neighbor_payoff_explanation}
\end{table}

\subsection{Contextual Properties}

The relevant context includes callee's location (home, party, meeting, library, and ER), relationship with caller (family, friend, colleague, stranger), and call urgency. 

\subsection{Types of Societies}
\label{sec:types_of_societies}

Agents optimize their norms based on a weighted sum of payoffs received by each stakeholder.
We define types of societies based on the generosity of their members in terms of the weight they place on the welfare (payoff) of their peers.

\begin{description}
    \item[Selfish] Members give weight only to their own payoff.
    \item[Pragmatic] Members give equal weight to everyone's payoff.
    \item[Considerate] Members give weight only to others' payoffs. 
    \item[Mixed] 25\% selfish, 25\% considerate, 50\% pragmatic. 
\end{description}

\section{Experiments and Results}
\label{sec:experiments}

To address our research questions, we run simulations of pragmatic, selfish, considerate, and mixed agent societies using three kinds of agents:

\begin{description}
    \item[Fixed] agents follow a fixed set of norms (Table~\ref{tab:fixed_agent_norm}). When norms conflict, they choose a random action. 
    \item[NSIGAs] evolve a set of explicit norms following our mechanism. When they violate a norm, they accept the sanction and use that feedback to guide learning. 
    \item[XSIGAs] go beyond NSIGAs by explaining their actions and issuing sanctions based on explanations received.
\end{description}

\begin{table}[t]
    \begin{tabular}{@{}c@{~~}c@{}}
         \begin{tabular}{@{} p{3.3cm} @{~} c@{}}
         \toprule
         \textbf{Location} & \textbf{Response}\\
         \midrule
         Emergency Room (ER) & Answer\\ 
         Home (H) & Answer\\ 
         Library (L) & Ignore\\ 
         Meeting (M) & Ignore\\ 
         Party (P) & Answer\\ 
         \bottomrule
        \end{tabular}
        
        & 
        
         \begin{tabular}{@{}l @{~} c @{~} c@{}} 
         \toprule
         \textbf{Circle} & \textbf{Casual} & \textbf{Urgent}\\
         \midrule
         Colleague & Answer & Answer\\ 
         Family & Answer & Answer\\ 
         Friend & Answer & Answer\\ 
         Stranger & Ignore & Answer\\ 
         \bottomrule\\
        \end{tabular}
    
    \end{tabular}
\caption{Norms followed by a fixed agent.}
\label{tab:fixed_agent_norm}
\end{table}

\subsection{Evaluation Metrics and Hypotheses}
\label{sec:eval_metrics}

We compute \emph{Social Experience}, \emph{Cohesion}, and \emph{Adoption}.

\emph{Social Experience} measures the degree of goal satisfaction delivered by an agent \citep{Ajmeri-IJCAI18-Poros}.
We compute it as the weighted aggregate of payoffs received by all of an agent's stakeholders as the result of its actions \citep{Ajmeri-IJCAI18-Poros}.
The weights depend on the nature of the agent (selfish, considerate, or pragmatic), as defined in Section~\ref{sec:types_of_societies}. 

\emph{Cohesion} measures the perception of norm compliance \citep{Ajmeri-IJCAI18-Poros}, and is the number of an agent's actions perceived as norm compliant by agents who are affected by them divided by the total number of interactions. 

\emph{Adoption} measures the percentage of agents in a society who comply with a particular norm. A norm is said to have emerged when adoption of the norm exceeds a threshold \citep{haynes2017engineering}. As in the literature, we consider 90\% adoption as the threshold \citep{delgado2002emergence}. 

We evaluate the following hypotheses: \textbf{H\fsub{1}} and \textbf{H\fsub{2}} relate to \textbf{RQ\fsub{G}}; \textbf{H\fsub{3}} and \textbf{H\fsub{4}} relate to \textbf{RQ\fsub{C}}; and \textbf{H\fsub{5}} relates to our overall research objective of studying the influence of explanations on norm emergence.
Each hypothesis compares the \emph{XSIGA} approach with the other approaches. We omit the corresponding null hypotheses for brevity.

\begin{description}
    \item[H\fsub{1}] XSIGAs give higher \emph{social experience} than \emph{Fixed} agents.
    \item[H\fsub{2}] XSIGAs give higher \emph{social experience} than \emph{NSIGAs}.
    \item[H\fsub{3}] XSIGAs give higher \emph{cohesion} than \emph{Fixed} agents.
    \item[H\fsub{4}] XSIGAs give higher \emph{cohesion} than \emph{NSIGAs}.
    \item[H\fsub{5}] XSIGAs give higher \emph{adoption} of norms than \emph{NSIGAs}.
\end{description}

We report results for each simulation run eight times for 10,000 timesteps. 
To evaluate these hypotheses, we conduct a paired t-test and  measure the effect size as Cohen's d. 

\subsection{Experiment: Pragmatic Agent Society}

\begin{figure*}[!htb]
\begin{subfigure}[b]{0.3\textwidth}
\begin{tikzpicture}
    \begin{axis}[
    title={},
    height=4.0cm,
    width=\columnwidth,
    xlabel={\small Time in 1,000 Steps},
    ylabel={\small Social Experience},
    xmin=-500,xmax=10500,
    ymin=0.05,ymax=2.05,
    xtick={0,2000,4000,6000,8000,10000},
    xticklabels={0,2,4,6,8,10},
    legend pos=south east,
    legend style={at={(1,0.4)}, anchor=east},
    legend columns=1,
    scaled ticks=false,
    ]
    \addplot +[mark size=1.2, draw=green!30!black, mark = triangle*] table [x=Step, y=Fixed, col sep=comma]
    {data/processed_data_for_tikz__Pragmatic_1_0_0_updated_base.csv};
    \addplot +[mark size=1.2, draw=tabblue, mark = o] table [x=Step, y=NSIGA, col sep=comma]
    {data/processed_data_for_tikz__Pragmatic_1_0_0_updated_base.csv};
    \addplot +[mark size=1.2, taborange, mark = square*] table [x=Step, y=XSIGA, col sep=comma]
    {data/processed_data_for_tikz__Pragmatic_1_0_0_updated_base.csv};
    \end{axis}
    \end{tikzpicture}
    \caption{Pragmatic society.}
    \label{fig:social_experience_1_0_0__1_15_17_sh2}
\end{subfigure}
\begin{subfigure}[b]{0.3\textwidth}
\begin{tikzpicture}
    \begin{axis}[
    title={},
    height=4.0cm,
    width=\columnwidth,
    xlabel={\small Time in 1,000 Steps},
    ylabel={\small Social Experience},
    xmin=-500,xmax=10500,
    ymin=0.05,ymax=2.05,
    xtick={0,2000,4000,6000,8000,10000},
    xticklabels={0,2,4,6,8,10},
    legend pos=south east,
    legend style={at={(1,0.4)}, anchor=east},
    legend columns=1,
    scaled ticks=false,
    ]
    \addplot +[mark size=1.2, draw=green!30!black, mark = triangle*] table [x=Step, y=Fixed, col sep=comma]
    {data/processed_data_for_tikz__Selfish_0_1_0_updated_base.csv};
    \addplot +[mark size=1.2, draw=tabblue, mark = o] table [x=Step, y=NSIGA, col sep=comma]
    {data/processed_data_for_tikz__Selfish_0_1_0_updated_base.csv};
    \addplot +[mark size=1.2, taborange, mark = square*] table [x=Step, y=XSIGA, col sep=comma]
    {data/processed_data_for_tikz__Selfish_0_1_0_updated_base.csv};
    \end{axis}
    \end{tikzpicture}
    \caption{Selfish society.}
    \label{fig:social_experience_0_1_0__1_15_17_sh2}
\end{subfigure}
\begin{subfigure}[b]{0.4\textwidth}
\begin{tikzpicture}
    \begin{axis}[
    title={},
    height=4.0cm,
    width=0.75\textwidth,
    xlabel={\small Time in 1,000 Steps},
    ylabel={\small Social Experience},
    xmin=-500,xmax=10500,
    ymin=0.05,ymax=2.05,
    xtick={0,2000,4000,6000,8000,10000},
    xticklabels={0,2,4,6,8,10},
    legend pos=south east,
    legend style={at={(1.5,0.4)}, anchor=east},
    legend columns=1,
    scaled ticks=false,
    ]
    \addplot +[mark size=1.2, draw=green!30!black, mark = triangle*] table [x=Step, y=Fixed, col sep=comma]
    {data/processed_data_for_tikz__Considerate_0_0_1_updated_base.csv};
    \addplot +[mark size=1.2, draw=tabblue, mark = o] table [x=Step, y=NSIGA, col sep=comma]
    {data/processed_data_for_tikz__Considerate_0_0_1_updated_base.csv};
    \addplot +[mark size=1.2, taborange, mark = square*] table [x=Step, y=XSIGA, col sep=comma]
    {data/processed_data_for_tikz__Considerate_0_0_1_updated_base.csv};
    \legend{\scriptsize Fixed, \scriptsize NSIGA, \scriptsize XSIGA}
    \end{axis}
    \end{tikzpicture}
    \caption{Considerate society.}
    \label{fig:social_experience_0_0_1__1_15_17_sh2}
\end{subfigure}

\caption{Comparing social experience yielded by Fixed, NSIGAs, and XSIGAs in pragmatic, selfish, and considerate agent societies.}
\end{figure*}
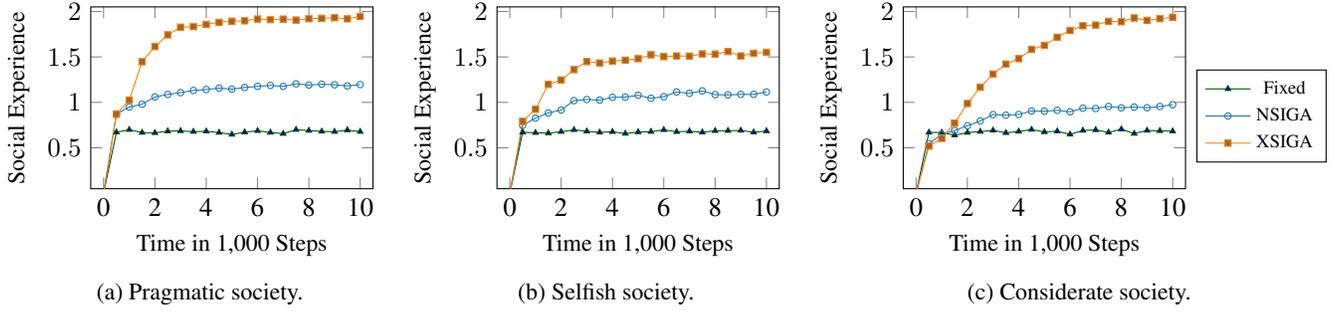

Table~\ref{tab:results_pragmatic_experience+cohesion} summarizes the social experience, cohesion, and adoption for the three agent types in a pragmatic society.
Figure~\ref{fig:social_experience_1_0_0__1_15_17_sh2} compares the social experience for Fixed, NSIGAs, and XSIGAs agents. 
We find the social experience, cohesion, and adoption yielded by XSIGAs to be better ($p<0.01$; d $>0.8$, indicating a large effect) than the two baselines.

\begin{table}[t]
    \begin{center}
    \npdecimalsign{.}
    \nprounddigits{2}
    \begin{tabular}{@{}l @{~~} n{1}{2} n{1}{2} n{2}{2} n{1}{2} n{2}{2} n{1}{2}@{}} 
     \toprule
     \multirow{2}{*}{\textbf{Agent}} & \multicolumn{2}{c}{\textbf{Experience}} & \multicolumn{2}{c}{\textbf{Cohesion}} & \multicolumn{2}{c}{\textbf{Adoption}}\\
     & {Mean} & {SD} & {Mean} & {SD} & {Mean} & {SD}\\
     \midrule
     Fixed & 0.683 & 0.01 & 27.061\% & 0.10 & {--} & {--}\\
     
     NSIGA & 1.206 & 0.01 & 55.646\% & 0.34 & 96.49\% & 0.1\\
     
     XSIGA & 1.936 & 0.01 & 88.812\% & 0.25 & 98.45\% & 0.22\\
     \bottomrule
    \end{tabular}
    \end{center}
\caption{Social experience and cohesion in a pragmatic society.}
\label{tab:results_pragmatic_experience+cohesion}
    
\end{table}

\begin{figure*}[!htb]
\begin{subfigure}[b]{0.33\textwidth}
    \centering
    \includegraphics[width=\columnwidth,height=3.5cm]{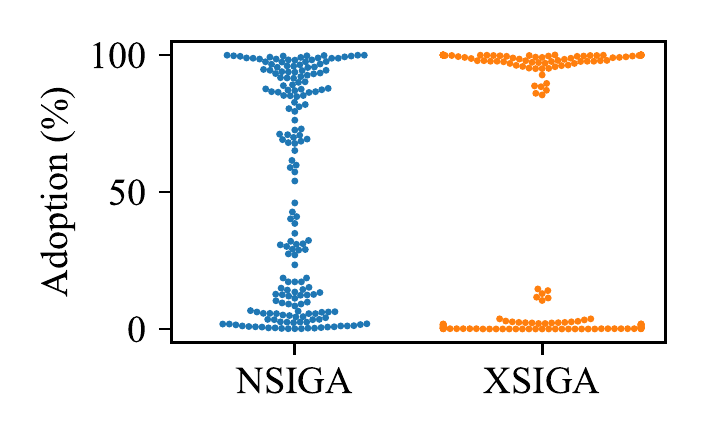}
    \caption{Adoption of norms: Pragmatic society.}
    \label{fig:norms_swarm_1_0_0__15_17_sh2}
\end{subfigure}
\begin{subfigure}[b]{0.33\textwidth}
    \centering
    \includegraphics[width=\columnwidth,height=3.5cm]{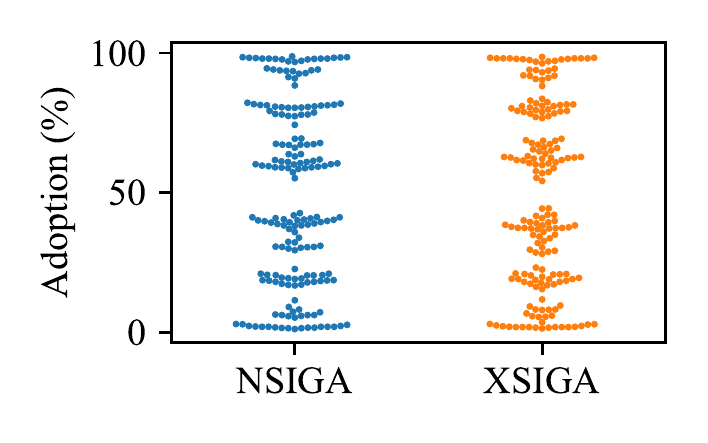}
    \caption{Adoption of norms: Selfish society.}
    \label{fig:norms_swarm_0_1_0__15_17_sh2}
\end{subfigure}
\begin{subfigure}[b]{0.33\textwidth}
    \centering
    \includegraphics[width=\columnwidth,height=3.5cm]{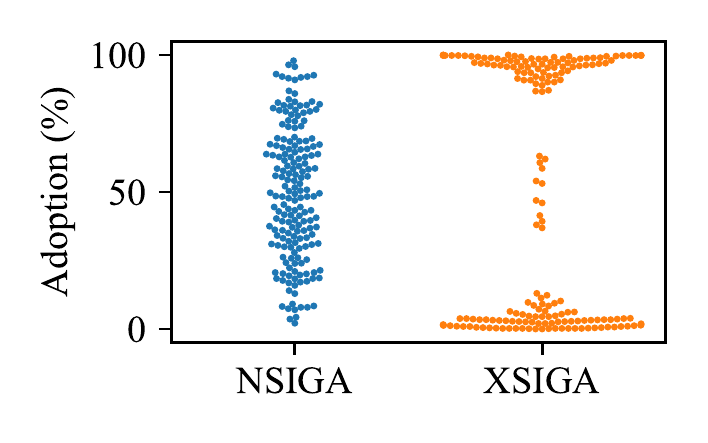}
    \caption{Adoption of norms: Considerate society.}
    \label{fig:norms_swarm_0_0_1__15_17_sh2}
\end{subfigure}
\caption{Comparing adoption of norms by Fixed, NSIGAs, and XSIGAs in pragmatic, selfish, and considerate agent societies.}
\label{fig:adoption-norms}
\end{figure*}

Figure~\ref{fig:norms_swarm_1_0_0__15_17_sh2} shows the adoption of norms for NSIGAs and XSIGAs (Fixed doesn't yield new norms). Each dot is a norm. Explanations help in identifying and promoting useful norms while discouraging the adoption of useless norms. As a result, XSIGAs produce a more extreme distribution.

Table~\ref{tab:emergent_norms_1_0_0_explanation_combined} shows the norms that emerge using SIGAs. The sole norm emerged with XSIGAs, \emph{always ring}, improves the payoffs overall.
This norm provides a better experience to the caller. The neighbor's experience depends on the acceptance or rejection of the explanation. The callee's experience can be negative, as for a casual call by a stranger.
NSIGAs get harsher penalties from neighbors for violating norms as they do not provide explanations.
As a result, the emerged norms are
more cautious: \emph{ring an urgent call}, \emph{ring a call by a known person}, and \emph{ring a call in an ER} because these provide sufficient positive payoffs to overcome possible negative neighbor reactions.
Whereas Figure~\ref{fig:norms_swarm_1_0_0__15_17_sh2} shows several norms with adoption above 90\%, Table~\ref{tab:emergent_norms_1_0_0_explanation_combined} lists only one emerged norm for XSIGAs, \emph{ring all calls}, because this norm is more general than the other emerged norms, such as \emph{ring urgent calls}.

\begin{table}[t]
    \centering
     \begin{tabular}{l l p{1.5cm}} 
     \toprule
     \textbf{Antecedent} & \textbf{Consequent} & \textbf{Adoption} \\
     \rowcolor{gray!20}\multicolumn{3}{c}{XSIGAs}\\
     true & ring & 91.2\%\\ 
     \rowcolor{gray!20}\multicolumn{3}{c}{NSIGAs}\\
     urgent = $\top$ & ring & 97.4\%\\ 
     
     callerRel = colleague & ring & 93.1\%\\ 
     
     callerRel = family & ring & 92.7\%\\ 
     
     calleeLoc = home & ring & 92.4\%\\ 
     
     callerRel = friend & ring & 92.2\%\\ 
     
     calleeLoc = ER & ring & 92.2\%\\ 
     \bottomrule
    \end{tabular}
\caption{Norms in a pragmatic society: XSIGAs and NSIGAs.}
\label{tab:emergent_norms_1_0_0_explanation_combined}
\end{table}

\subsection{Experiment: Selfish Agent Society}

Table~\ref{tab:results_selfish_social_experience+cohesion} summarizes the social experience, cohesion, and adoption yielded by the three agent types in a selfish society.
As Figure~\ref{fig:social_experience_0_1_0__1_15_17_sh2} shows, we find the social experience and cohesion for XSIGAs to be better ($p<0.01$; d $>0.8$, indicating a large effect) than the baselines, and reject the null hypotheses corresponding to H\fsub{1}--H\fsub{4}. For adoption, however, the difference in mean values is not statistically significant and we fail to reject the null hypothesis corresponding to H\fsub{5}.

\begin{table}[t]
    \begin{center}
    \npdecimalsign{.}
    \nprounddigits{2}
    \begin{tabular}{@{}l @{~~} n{1}{2} n{1}{2} n{2}{2} n{1}{2} n{2}{2} n{1}{2}@{}} 
     \toprule
     \multirow{2}{*}{\textbf{Agent}} & \multicolumn{2}{c}{\textbf{Experience}} & \multicolumn{2}{c}{\textbf{Cohesion}} & \multicolumn{2}{c}{\textbf{Adoption}}\\
     & {Mean} & {SD} & {Mean} & {SD} & {Mean} & {SD}\\
     \midrule
     Fixed & 0.68 & 0.02 & 27.09\% & 0.13 & {--} & {--}\\
     
     NSIGA & 1.10 & 0.01 & 49.20\% & 0.13 & 96.07\% & 0.22\\
     
     XSIGA & 1.55 & 0.01 & 68.46\% & 0.23 & 95.94\% & 0.31\\
     \bottomrule
    \end{tabular}
    \end{center}
\caption{Social experience and cohesion in a selfish society.}
\label{tab:results_selfish_social_experience+cohesion}
\end{table}

Figure~\ref{fig:norms_swarm_0_1_0__15_17_sh2} shows the norm adoption for the SIGA approaches. Providing explanations has a lower effect on emergence in a selfish society than pragmatic or considerate societies. 
Mean adoption among emerged norms is similar for XSIGAs and NSIGAs because selfish agents don't give weight to anyone's payoff other than their own and do not give value to responses from other agents. Hence, selfish XSIGAs learn norms without regard to how others evaluate their explanations though, as XIGAs, they provide and evaluate explanations. 
Providing an explanation is better in terms of social experience and cohesion because the neighbors are also selfish, and would accept an explanation when they would have done the same thing as the callee in the same context, even if it is a selfish action.
This acceptance of explanations leads to better social experience and cohesion, even if it doesn't affect the exact norms that emerge.

Table~\ref{tab:emergent_norms_0_1_0_explanation_combined} shows the norms that have emerged.
The norms emerged for XSIGAs and NSIGAs are similar to each other, in line with our expectation that norms emerging in a selfish society won't be affected by providing explanations. The emerged norms are: \emph{pick up an urgent call} and \emph{always ignore a casual call from a stranger}. This is because a selfish agent does not value the payoff of neighbors or callers. It ignores a casual call from a stranger even in a nonrestrictive location like ER where the neighbors expect it to ring a call.

\begin{table}[t]
    \centering
     \begin{tabular}{@{}l l @{~~} p{1.4cm}@{}} 
     \toprule
     \textbf{Antecedent} & \textbf{Consequent} & \textbf{Adoption} \\
     \rowcolor{gray!20}\multicolumn{3}{c}{XSIGAs}\\
     urgent = $\top$ & ring & 91.4\%\\ 
     urgent = $\bot$ $\land$ callerRel = stranger & ignore & 90.2\%\\ 
     \rowcolor{gray!20}\multicolumn{3}{c}{NSIGAs}\\
     urgent = $\top$ & ring & 91.4\%\\ 
     urgent = $\bot$ $\land$ callerRel = stranger & ignore & 90.2\%\\ 
     \bottomrule
    \end{tabular}
\caption{Norms in a selfish society: XSIGAs and NSIGAs.}
\label{tab:emergent_norms_0_1_0_explanation_combined}
\end{table}

\subsection{Experiment: Considerate Agent Society}

Table~\ref{tab:results_considerate_social_experience+cohesion} summarizes the social experience, cohesion, and adoption for the three agent types in a considerate society.

\begin{table}[t]
    \begin{center}
    \npdecimalsign{.}
    \nprounddigits{2}
    \begin{tabular}{@{}l n{1}{2} n{1}{2} n{2}{2} n{1}{2} n{2}{2} n{1}{2}@{}} 
     \toprule
     \multirow{2}{*}{\textbf{Agent}} & \multicolumn{2}{c}{\textbf{Experience}} & \multicolumn{2}{c}{\textbf{Cohesion}} & \multicolumn{2}{c}{\textbf{Adoption}}\\
     & {Mean} & {SD} & {Mean} & {SD} & {Mean} & {SD}\\
     \midrule
     Fixed & 0.69 & 0.01 & 27.11\% & 0.14 & {--} & {--}\\
     
     NSIGA & 0.97 & 0.02 & 69.72\% & 0.24 & 93.70\% & 0.23\\
     
     XSIGA & 1.93 & 0.01 & 77.48\% & 0.63 & 97.18\% & 0.3\\
     \bottomrule
    \end{tabular}
    \end{center}
\caption{Social experience, cohesion in a considerate society.}
\label{tab:results_considerate_social_experience+cohesion}
\end{table}

Figure~\ref{fig:social_experience_0_0_1__1_15_17_sh2} compares the social experience plots for Fixed, NSIGAs, and XSIGAs agents in a considerate society. 
We find the social experience, cohesion, and adoption yielded by XSIGAs to be better ($p<0.01$; d $>0.8$, indicating a large effect) than the two baselines, and thus reject the null hypotheses corresponding to H\fsub{1}--H\fsub{5}.
Figure~\ref{fig:norms_swarm_0_0_1__15_17_sh2} shows the adoption of norms for different approaches in a considerate society. As in a pragmatic society, providing explanations had a polarizing influence on adoption. This helps in the emergence of norms by increasing the adoption of the emerged norms.

Table~\ref{tab:emergent_norms_0_0_1_explanation_combined} shows explicit norms that emerge using SIGAs. For XSIGAs, the effective norm is to \emph{ring in all cases} for similar reasons as for pragmatic agents. We observe that even within the norm, a more specialized version like \emph{ring an urgent call} has higher adoption than \emph{ring a casual call}. For NSIGAs, the agent needs to be more careful with the neighbors' payoffs, as for pragmatic agents. As a result, the most adopted norms are based on location and urgency.

\begin{table}[t]
    \centering
     \begin{tabular}{@{}l l @{~~} p{1.4cm}@{}} 
     \toprule
     \textbf{Antecedent} & \textbf{Consequent} & \textbf{Adoption} \\
     \rowcolor{gray!20}\multicolumn{3}{c}{XSIGAs}\\
     urgent = $\top$ & ring & 97.4\%\\ 
     urgent = $\bot$ & ring & 92.2\%\\ 
     \rowcolor{gray!20}\multicolumn{3}{c}{NSIGAs}\\
     urgent = $\top$ $\land$ calleeLoc = home & ring & 92.0\%\\ 
     
     urgent = $\top$ $\land$ calleeLoc = ER & ring & 92.3\%\\ 
     
     urgent = $\bot$ $\land$ callerRel = stranger & ignore & 90.2\%\\ 
     \bottomrule
    \end{tabular}
\caption{Norms in a considerate society: XSIGAs and NSIGAs.}
\label{tab:emergent_norms_0_0_1_explanation_combined}
\end{table}

\section{Discussion}
\label{sec:discussions}

We find that societies composed of XSIGAs have better social experience and cohesion than the baselines and that providing and evaluating explanations leads to better adoption of emerged norms, except for a selfish society.

\citet{mosca2021elvira} generate explanations in multiuser privacy scenarios for evaluation by a human. In contrast, our explanations are communicated between agents for norm emergence. 
\citet{hind2019ted} apply machine learning to jointly predict actions and explanations. In contrast, our explanations are formed of the norms that explain an action.

\citet{Morales-AAMAS14-Minimality} study centralized norm emergence where conflict scenarios are recognized and norms adapted to handle the conflicts. But, here, agents themselves create the norms.
\citet{Morales2018Offline} synthesize  norms using an offline, centralized mechanism to evolve norms aligned with goals, whereas here norms evolve in an online, decentralized manner based on sanctions.
\citepos{Mashayekhi+2022-Cha}
decentralized norm emergence framework is driven by conflict detection with norms created to avoid conflicts. But, here, SIGAs are driven by sanctions and sharing explanations.

\citet{DellAnna-JAAMAS20-Runtime} use Bayesian Networks to revise sanctions associated with enforced norms. In contrast, we revise the norms themselves while the sanctions stay the same.
\citet{Hao+TAAS18-efficient} propose heuristic collective learning frameworks to learn norms as best responses to all states, whereas we apply genetic exploration to evolve the antecedents used in the norms instead of learning for all states.

\citet{Ajmeri-IJCAI18-Poros} provide an implicit norm system that learns actions to be taken in different contexts, whereas our work is an explicit norm system that explicitly learns and reasons about norms. They share not explanations but the entire context in which the decision was made to violate the norm. This context is evaluated by other agents to decide if they would have done the same action  as a basis for sanctioning. 
In contrast, we share explicit norms that influence a decision. 

This work suggests important and interesting extensions. One, to generate explanations that trade off privacy and specificity---revealing more information may yield clarity at the loss of privacy. Two, to apply a domain ontology to learn norms with complex boundaries reflecting the subtleties of real-life norms. Three, to enhance SIGAs to incorporate values and value preferences, ethics, and fairness to guide explicit norm emergence \citep{AAMAS-20:Elessar,AAMAS-20:Blue-Sky,Santos-AAAI19-FairnessEvolution,Serramia+18:values,Woodgate+Ajmeri-AAMAS22-BlueSky}.

\section*{Acknowledgments}
RA is now with Google. NA thanks the University of Bristol for support.
MPS thanks the US National Science Foundation (grant IIS-2116751) for support.

\DeclareRobustCommand{\nUmErAL}[1]{#1}\DeclareRobustCommand{\nAmE}[3]{#3}

\appendix

\section{Code Availability} 

Code for the ringer simulation and the generating plots is available here: \url{https://github.com/niravajmeri/SIGA-IJCAI2022/}.

\section{Hyperparameters}
The \emph{XCS} algorithm used in the SIGA has many parameters that controls its behavior. We have touched upon some of them like learning rate, size of the rule population, fitness scaling factor etc. We summarize the hyperparameters we have used in the Table~\ref{tab:hyperparameters}. We used the standard settings for these parameters \citep{urbanowicz2017introduction}, except population size, which is set at 30 due to the smaller dimensionality of our problem. 

\begin{table}[ht!]
    \begin{center}
    \caption{Hyperparameters for Our Method.}
    \label{tab:hyperparameters}
     \begin{tabular}{l S} 
     \toprule
     \textbf{Parameter} & \textbf{Value}\\
     \midrule
     Max Population size & 30\\ 
     
     Don't care probability & 0.3\\ 
     
     Accuracy threshold & 0.01\\ 
     
     Fitness exponent & 5\\ 
     
     Learning rate & 0.1\\
     
     GA threshold & 25\\
     
     Mutation probability & 0.4\\
     
     Crossover probability & 0.8\\
     
     Experience threshold for deletion & 20\\
     
     Experience threshold for subsumption & 20\\
     
     Fitness falloff & 0.1\\
     \bottomrule
    \end{tabular}
    \end{center}
\end{table}

\section{Threats to Validity}
First, our evaluation is based on manually created payoffs. Getting correct valuation of user experience from actual users is difficult. We mitigate this threat by adapting data from the literature. Appendix~\ref{chap:res_diff_sets_of_payoffs} (in supplement) demonstrates the stability of our results on other payoffs.
Second, user attitudes differ in the relative value placed on the welfare of other people. We represent societies with different types of attitudes to show the robustness of our approach. In addition to results in pragmatic, selfish, and considerate societies, Appendix~\ref{app:mixed-society} (in supplement) demonstrates the results in a mixed society. 
%

\section{Experiment with Mixed Agent Society}
\label{app:mixed-society}

\begin{figure}[!htb]
    \centering
    \includegraphics[width=\columnwidth]{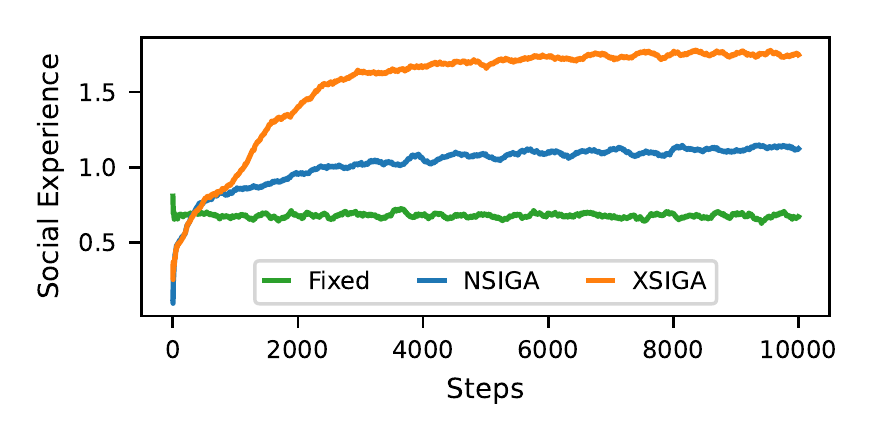}
%
    \includegraphics[width=\columnwidth]{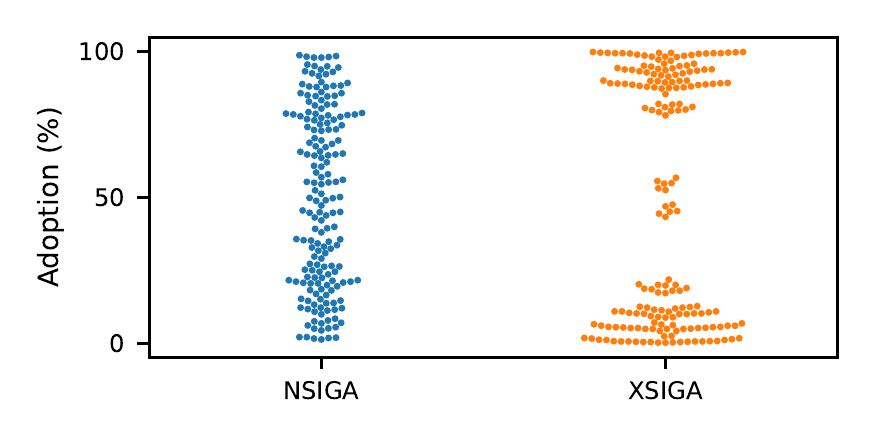}
\caption{Social experience and norms adoption in a mixed society.}
\label{fig:mixed-society}
\end{figure}

Table~\ref{tab:results_mixed_social_experience+cohesion} summarizes the social experience, cohesion and adoption metrics yielded by the three agent types in a mixed society.
Figure~\ref{fig:mixed-society} compares the social experience plots for Fixed agents, NSIGAs, and XSIGAs in a mixed society. 
We find the social experience, cohesion, and adoption yielded by XSIGAs to be better ($p<0.01$; d$>0.8$, indicating large effect) than the two baselines, and thus reject the null hypotheses corresponding to H\fsub{1}--H\fsub{5}.

\begin{table}[ht!]
    \caption{Social experience and cohesion in mixed society.}
    \label{tab:results_mixed_social_experience+cohesion}
    \begin{center}
    \npdecimalsign{.}
    \nprounddigits{2}
    \begin{tabular}{@{}l n{1}{2} n{1}{2} n{2}{2} n{1}{2} n{2}{2} n{1}{2}@{}} 
     \toprule
     \multirow{2}{*}{\textbf{Agent}} & \multicolumn{2}{c}{\textbf{Experience}} & \multicolumn{2}{c}{\textbf{Cohesion}} & \multicolumn{2}{c}{\textbf{Adoption}}\\
     & {Mean} & {SD} & {Mean} & {SD} & {Mean} & {SD}\\
     \midrule
     Fixed & 0.67 & 0.01 & 26.99\% & 0.11 & {--} & {--}\\
     
     NSIGA & 1.12 & 0.01 & 57.55\% & 0.22 & 95.09\% & 0.18\\
     
     XSIGA & 1.76 & 0.01 & 77.69\% & 0.29 & 95.79\% & 0.25\\
     \bottomrule
    \end{tabular}
    \end{center}
\end{table}

Figure~\ref{fig:mixed-society} shows adoption of norms for different approaches in mixed society. Same polarizing effect as pragmatic and considerate societies, can be observed in this case as well.


Tables~\ref{tab:emergent_norms_2_1_1_explanation_combined} shows explicit norms that have emerged using SIGAs. For a society of XSIGAs, the emerged norm is to ring urgent calls and calls from known people. This is more specific than the pragmatic society norm of always ringing because there are also selfish agents who prefer to ignore casual calls from strangers. For the society of NSIGAs, we see that callee's location is important in adopted norms because of neighbor payoffs, as in the other societies.

\begin{table}[htb!]
    \caption{Norms in Mixed Society, XSIGAs and NSIGAs.}
    \label{tab:emergent_norms_2_1_1_explanation_combined}
    \centering
     \begin{tabular}{@{}l @{~~} l @{~~}p{1.1cm}@{~~}} 
     \toprule
     \textbf{Antecedent} & \textbf{Consequent} & \textbf{Adoption} \\
     \rowcolor{gray!20}\multicolumn{3}{c}{XSIGAs}\\
%
     urgent = true & ring & 96.8\%\\ 
     
     callerRel = colleague & ring & 92.0\%\\ 
     
     callerRel = friend & ring & 91.8\%\\ 
     
     callerRel = family & ring & 91.4\%\\ 
     \rowcolor{gray!20}\multicolumn{3}{c}{NSIGAs}\\
     urgent = true $\land$ calleeLoc = home & ring & 95.5\%\\ 
     
     urgent = true $\land$ calleeLoc = ER & ring & 95.0\%\\ 
     
     urgent = true $\land$ callerRel = friend & ring & 94.9\%\\ 
     
     urgent = true $\land$ callerRel = family & ring & 94.5\%\\ 
     
     urgent = true $\land$ callerRel = colleague & ring & 93.8\%\\ 
     \bottomrule
    \end{tabular}
\end{table}

\section{Results with Statistical Analysis}

\subsection{Pragmatic Society}

Tables~\ref{tab:results_pragmatic_social_experience}, \ref{tab:results_pragmatic_social_cohesion} and \ref{tab:results_pragmatic_adoption} list the experience, cohesion and adoption offered by Fixed, NSIGA, and XSIGA in a pragmatic society. The tables include $p$-values from t-test and Cohen's d for effect size. 

\begin{table}[!htb]
    \caption{Social experience experiment in pragmatic society.}
    \label{tab:results_pragmatic_social_experience}
    \begin{center}
     \begin{tabular}{@{}lS S S@{}} 
     \toprule
     \textbf{Agent Type} & \textbf{Experience} & \textbf{$p$-value} & \textbf{Cohen's d}\\
     \midrule
     Fixed & 0.683 & <0.0001 & 104.427\\
     
     NSIGA & 1.206 & <0.0001 & 69.380\\
     
     XSIGA & 1.936 & {--} & {--}\\
     \bottomrule
    \end{tabular}
    \end{center}
\end{table}

\begin{table}[!htb]
    \caption{Cohesion in a pragmatic society.}
    \label{tab:results_pragmatic_social_cohesion}
    \begin{center}
     \begin{tabular}{@{}lS S S@{}} 
     \toprule
     \textbf{Agent Type} & \textbf{Cohesion} &\textbf{$p$-value} & \textbf{Cohen's d}\\
     \midrule
     Fixed & 27.061\% & <0.0001 & 322.159\\
     
     NSIGA & 55.646\% & <0.0001 & 111.463\\
     
     XSIGA & 88.812\% & {--} & {--}\\
     \bottomrule
    \end{tabular}
    \end{center}
\end{table}

\begin{table}[!htb]
    \caption{Adoption in a pragmatic society.}
    \label{tab:results_pragmatic_adoption}
    \begin{center}
     \begin{tabular}{@{}lS S S@{}} 
     \toprule
     \textbf{Agent Type} & \textbf{Adoption} &\textbf{$p$-value} & \textbf{Cohen's d}\\
     \midrule
     NSIGA & 96.49\% & <0.001 & 11.62\\
     
     XSIGA & 98.71\% & {--} & {--}\\
     \bottomrule
    \end{tabular}
    \end{center}
\end{table}

\subsection{Selfish Society}

Tables~\ref{tab:results_selfish_social_experience}, \ref{tab:results_selfish_social_cohesion} and \ref{tab:results_selfish_adoption} list the experience, cohesion and adoption offered by Fixed, NSIGA, and XSIGA in a selfish society. The tables include $p$-values from t-test and Cohen's d for effect size. 

\begin{table}[!htb]
    \caption{Social experience in selfish society.}
    \label{tab:results_selfish_social_experience}
    \begin{center}
     \begin{tabular}{@{}lS S S@{}} 
     \toprule
     \textbf{Agent Type} & \textbf{Experience} & \textbf{$p$-value} & \textbf{Cohen's d}\\
     \midrule
     Fixed & 0.683 & <0.0001 & 50.468\\
     
     NSIGA & 1.100 & <0.0001 & 37.551\\
     
     XSIGA & 1.547 & {--} & {--}\\
     \bottomrule
    \end{tabular}
    \end{center}
\end{table}

\begin{table}[!htb]
    \begin{center}
    \caption{Cohesion in a selfish society.}
    \label{tab:results_selfish_social_cohesion}
     \begin{tabular}{@{}lS S S@{}} 
     \toprule
     \textbf{Agent Type} & \textbf{Cohesion} & \textbf{$p$-value} & \textbf{Cohen's d}\\
     \midrule
     Fixed & 27.085\% & <0.0001 & 218.574\\
     
     NSIGA & 49.195\% & <0.0001 & 101.538\\
     
     XSIGA & 68.456\% & {--} & {--}\\
     \bottomrule
    \end{tabular}
    \end{center}
\end{table}

\begin{table}[!htb]
    \caption{Adoption in a selfish society.}
    \label{tab:results_selfish_adoption}
    \begin{center}
     \begin{tabular}{@{}lS S S@{}} 
     \toprule
     \textbf{Agent Type} & \textbf{Adoption} &\textbf{$p$-value} & \textbf{Cohen's d}\\
     \midrule
     NSIGA & 96.06\% & 0.66 & 0.45\\
     
     XSIGA & 95.94\% & {--} & {--}\\
     \bottomrule
    \end{tabular}
    \end{center}
\end{table}

\subsection{Considerate Society}

Tables~\ref{tab:results_considerate_social_experience}, \ref{tab:results_considerate_social_cohesion} and \ref{tab:results_considerate_adoption} list the experience, cohesion and adoption offered by Fixed, NSIGA, and XSIGA in a considerate society. The tables include $p$-values from t-test and Cohen's d for effect size. 

\begin{table}[!htb]
    \caption{Social experience in considerate society.}
    \label{tab:results_considerate_social_experience}
    \begin{center}
          \begin{tabular}{@{}lS S S@{}} 
     \toprule
     \textbf{Agent Type} & \textbf{Experience} & \textbf{$p$-value} & \textbf{Cohen's d}\\
     \midrule
     Fixed & 0.687 & <0.0001 & 160.394\\
     
     NSIGA & 0.973 & <0.0001 & 76.529\\
     
     XSIGA & 1.933 & {--} & {--}\\
     \bottomrule
    \end{tabular}
    \end{center}
\end{table}

\begin{table}[!htb]
    \caption{Cohesion in a considerate society.}
    \label{tab:results_considerate_social_cohesion}
    \begin{center}
     \begin{tabular}{@{}lS S S@{}} 
     \toprule
     \textbf{Agent Type} & \textbf{Cohesion} & \textbf{$p$-value} & \textbf{Cohen's d}\\
     \midrule
     Fixed & 27.114\% & <0.0001 & 109.540\\
     
     NSIGA & 69.720\% & <0.0001 & 16.179\\
     
     XSIGA & 77.482\% & {--} & {--}\\
     \bottomrule
    \end{tabular}
    \end{center}
\end{table}

\begin{table}[!htb]
    \caption{Adoption in a considerate society.}
    \label{tab:results_considerate_adoption}
    \begin{center}
     \begin{tabular}{@{}lS S S@{}} 
     \toprule
     \textbf{Agent Type} & \textbf{Adoption} &\textbf{$p$-value} & \textbf{Cohen's d}\\
     \midrule
     NSIGA & 93.70\% & <0.001 & 13.18\\
     
     XSIGA & 97.18\% & {--} & {--}\\
     \bottomrule
    \end{tabular}
    \end{center}
\end{table}

\subsection{Mixed Society}

Tables~\ref{tab:results_mixed_social_experience}, \ref{tab:results_mixed_social_cohesion} and \ref{tab:results_mixed_adoption} list the experience, cohesion and adoption offered by Fixed, NSIGA, and XSIGA in a mixed society. The tables include $p$-values from t-test and Cohen's d for effect size. 

\begin{table}[!htb]
    \caption{Social experience in mixed society.}
    \label{tab:results_mixed_social_experience}
    \begin{center}
     \begin{tabular}{@{}lS S S@{}} 
     \toprule
     \textbf{Agent Type} & \textbf{Experience} & \textbf{$p$-value} & \textbf{Cohen's d}\\
     \midrule
     Fixed & 0.670 & <0.0001 & 86.769\\
     
     NSIGA & 1.115 & <0.0001 & 52.162\\
     
     XSIGA & 1.759 & {--} & {--}\\
     \bottomrule
    \end{tabular}
    \end{center}
\end{table}

\begin{table}[!htb]
    \caption{Cohesion in a mixed society.}
    \label{tab:results_mixed_social_cohesion}
    \begin{center}
     \begin{tabular}{@{}lS S S@{}} 
     \toprule
     \textbf{Agent Type} & \textbf{Cohesion} & \textbf{$p$-value} & \textbf{Cohen's d}\\
     \midrule
     Fixed & 26.990\% & <0.0001 & 231.032\\
     
     NSIGA & 57.549\% & <0.0001 & 77.492\\
     
     XSIGA & 77.689\% & {--} & {--}\\
     \bottomrule
    \end{tabular}
    \end{center}
\end{table}

\begin{table}[!htb]
    \caption{Adoption in a mixed society.}
    \label{tab:results_mixed_adoption}
    \begin{center}
     \begin{tabular}{@{}lS S S@{}} 
     \toprule
     \textbf{Agent Type} & \textbf{Adoption} &\textbf{$p$-value} & \textbf{Cohen's d}\\
     \midrule
     NSIGA & 95.09\% & <0.001 & 3.21\\
     
     XSIGA & 95.79\% & {--} & {--}\\
     \bottomrule
    \end{tabular}
    \end{center}
\end{table}

\section{Comparison with Poros}
\label{chap:comparison_poros}

We do not include Poros \citep{Ajmeri-IJCAI18-Poros} as a baseline in the main paper because as our aim in this work is to study norm emergence in explicit norm systems and how providing explanations impact the emergence of norms. Poros, on the other hand, is an implicit norm system that does not reason about explicit norms. A comparison with Poros does not directly help us in answering our research questions. However, for completeness, we provide a comparison here.

Agents in \citet{Ajmeri-IJCAI18-Poros}'s work are implicit norm agents that use linear regression on past data to predict the expected payoff of each action and then choose the action with the best payoff. 

\subsection{Social Experience}
\begin{figure}[htb]
    \centering
    \includegraphics[width=\columnwidth]{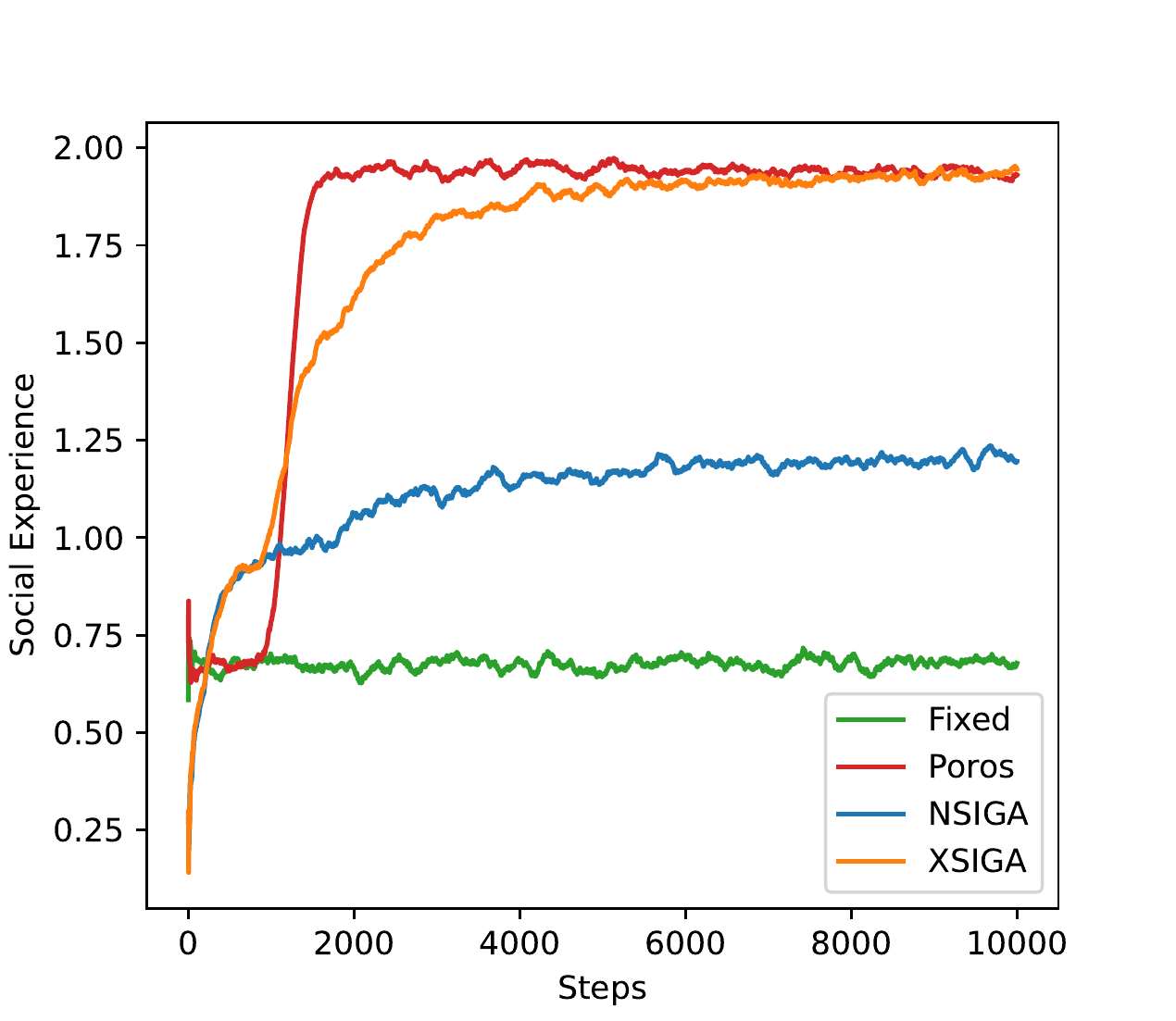}
    \caption[Social Experience in Pragmatic Society.]{Social Experience in Pragmatic Society.}
    \label{fig:social_experience_1_0_0__1_3_15_17_sh2}
\end{figure}

Poros agents and XSIGA provide similar social experience as can be seen in Figure~\ref{fig:social_experience_1_0_0__1_3_15_17_sh2}.

\subsection{Cohesion}
Poros agents and XSIGA also have similar cohesion after 3000 steps, 74.25\% and 73.77\% respectively.

\subsection{Decision Time}

The time taken to choose an action increases with the amount of past data for Poros agents, whereas the time stays constant for SIGAs as we don't store past data. Table~\ref{tab:decision_time} shows the average time taken to decide the action after 3000 steps of the simulation. We can see that SIGAs are about 50 times faster than Poros agents, after 3000 steps of simulation.

\begin{table}[ht!]
    \caption{Time to decide action.}
    \label{tab:decision_time}
    \begin{center}
     \begin{tabular}{l S} 
     \toprule
     \textbf{Agent Type} & \textbf{Time to decide (ms)}\\
     \midrule
     Poros & 0.4355\\
     
     NSIGA & 0.0081\\
     
     XSIGA & 0.0079\\
     \bottomrule
    \end{tabular}
    \end{center}
\end{table}

\subsection{Privacy}

We define a metric called \emph{Privacy Score} to compare the amount of information revealed by the agents. We define privacy score as the proportion of contextual properties not revealed by the agent.
\begin{equation*}
    \mbox{Privacy Score} = \frac{\mbox{\# of contextual properties not revealed by agent}}{\mbox{\# of contextual properties}}
\end{equation*}
Poros agents reveal the entire context and thus have the privacy score of 0, whereas XSIGA have an average privacy score of 0.197. Thus, SIGAs reveal 19.7\% less information than Poros agents.

\subsection{Conclusion}

XSIGA provide similar social experience and cohesion as Poros agents, but are more private and much faster. In addition to that, they are explainable as the behavior is driven by explicit norms that can be manually inspected, if needed, whereas Poros agents are driven by just regression to choose highest payoff value and do not reason about norms.

\section{Results on different sets of payoffs}
\label{chap:res_diff_sets_of_payoffs}

\subsection{Poros payoffs}
In this section, we present our results using the exact payoffs from \citet{Ajmeri-IJCAI18-Poros}. 
The payoffs are in Tables~\ref{tab:app_poros_callee_payoff}, \ref{tab:app_poros_caller_payoff}, \ref{tab:app_poros_neighbor_payoff_fixed} and~\ref{tab:app_poros_neighbor_payoff_explanation}.

\begin{table}[ht!]
    \centering
    \caption{Callee payoff based on call urgency and caller relationship.}
    \label{tab:app_poros_callee_payoff}
     \begin{tabular}{p{3cm}p{1.5cm}SS} 
     \toprule
     \textbf{Caller Relationship} & \textbf{Callee's Action} & \textbf{Casual} & \textbf{Urgent}\\
     \midrule
     \multirow{2}{*}{\parbox{3cm}{Family, Friend or Colleague}} & Answer & 0.50 & 1.00\\ 
     & Ignore & 0.00 & -0.50\\ 
     \multirow{2}{*}{Stranger} & Answer & 0.00 & 0.50\\ 
     & Ignore & 0.25 & -0.25\\ 
     \bottomrule
    \end{tabular}
\end{table}

\begin{table}[ht!]
    \centering
    \caption{Caller payoff based on urgency of the call.}
    \label{tab:app_poros_caller_payoff}
     \begin{tabular}{lSS} 
     \toprule
     \textbf{Callee's Action} & \textbf{Casual} & \textbf{Urgent}\\
     \midrule
     Answer & 0.5 & 1.00\\ 
     Ignore & -0.5 & -1.00\\ 
     \bottomrule
    \end{tabular}
\end{table}

\begin{table}[ht!]
    \centering
    \caption{Neighbor payoff based on location of the call}
    \label{tab:app_poros_neighbor_payoff_fixed}
     \begin{tabular}{@{}p{1.5cm} S@{~} S@{~} S@{~} S@{~} S@{~}} 
     \toprule
     \textbf{Callee's Action} & \textbf{ER} & \textbf{H} & \textbf{L} & \textbf{M} & \textbf{P}\\
     \midrule
     Answer & 1.00 & 0.67 & -1.00 & -1.00 & -0.33\\ 
     Ignore & -1.00 & -0.33 & 1.00 & 1.00 & 0.67\\ 
     \bottomrule
    \end{tabular}
\end{table}

\begin{table}[ht!]
    \centering
    \caption{Neighbor payoff based on evaluation of explanation}
    \label{tab:app_poros_neighbor_payoff_explanation}
     \begin{tabular}{@{}l l S@{~} S@{~} S@{~} S@{~} S@{~}} 
     \toprule
     \parbox{1.0cm}{\textbf{Callee Action}} & \parbox{1.2cm}{\textbf{Neighbor Expects}} & \textbf{ER} & \textbf{H} & \textbf{L} & \textbf{M} & \textbf{P}\\
     \midrule
     Answer & Answer & 1.00 & 0.67 & 1.00 & 1.00 & 0.67\\ 
     Answer & Ignore & -1.00 & -0.33 & -1.00 & -1.00 & -0.33\\ 
     Ignore & Answer & -1.00 & -0.33 & -1.00 & -1.00 & -0.33\\ 
     Ignore & Ignore & 1.00 & 0.67 & 1.00 & 1.00 & 0.67\\ 
     \bottomrule
    \end{tabular}
\end{table}

\begin{figure*}[!ht]
    \centering
\begin{subfigure}[b]{0.45\textwidth}
    \includegraphics[width=\columnwidth]{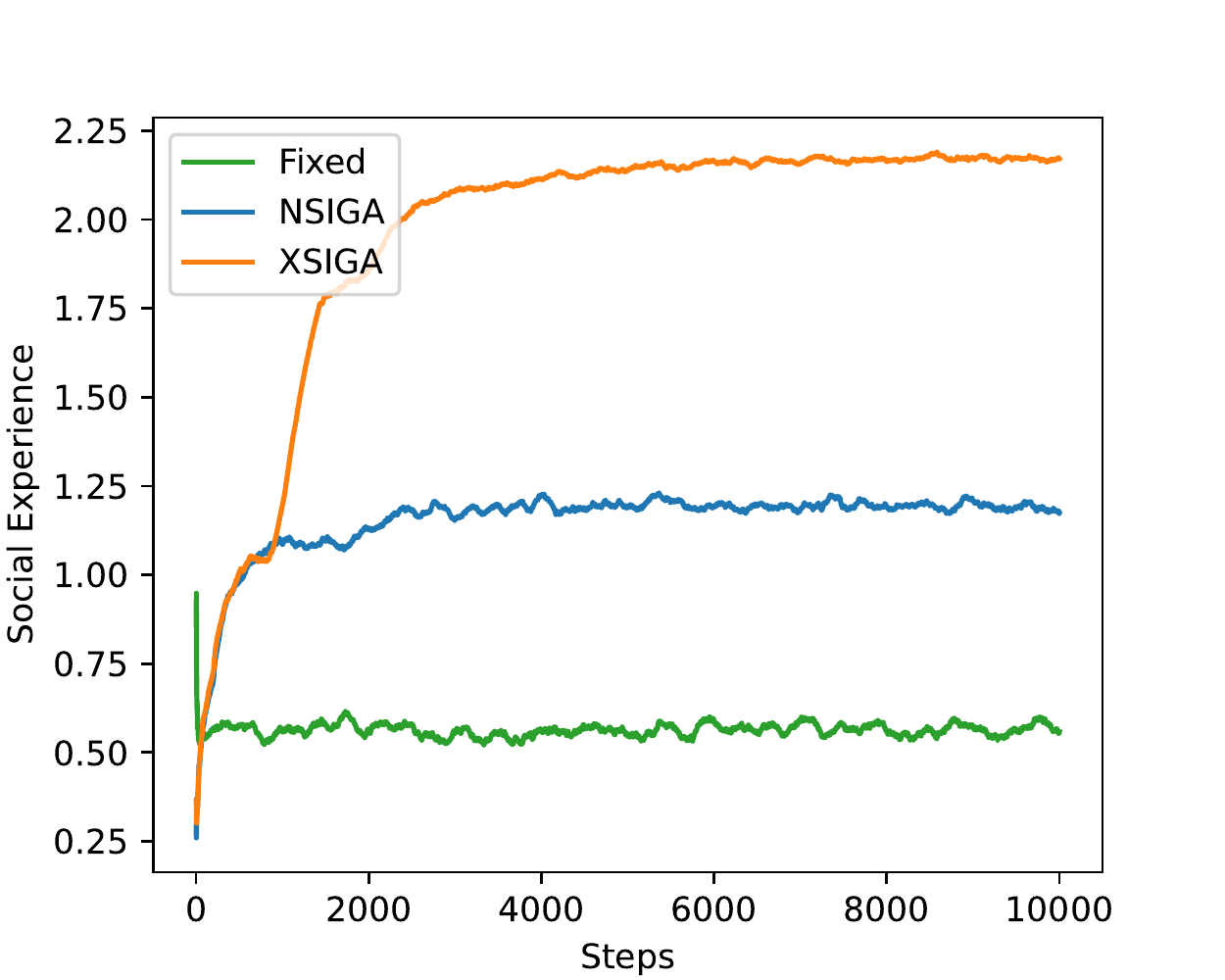}
    \caption[Social Experience in Pragmatic Society.]{Social Experience in Pragmatic Society.}
    \label{fig:social_experience_1_0_0__1_15_17_sh1}
\end{subfigure}
\begin{subfigure}[b]{0.45\textwidth}
    \includegraphics[width=\columnwidth]{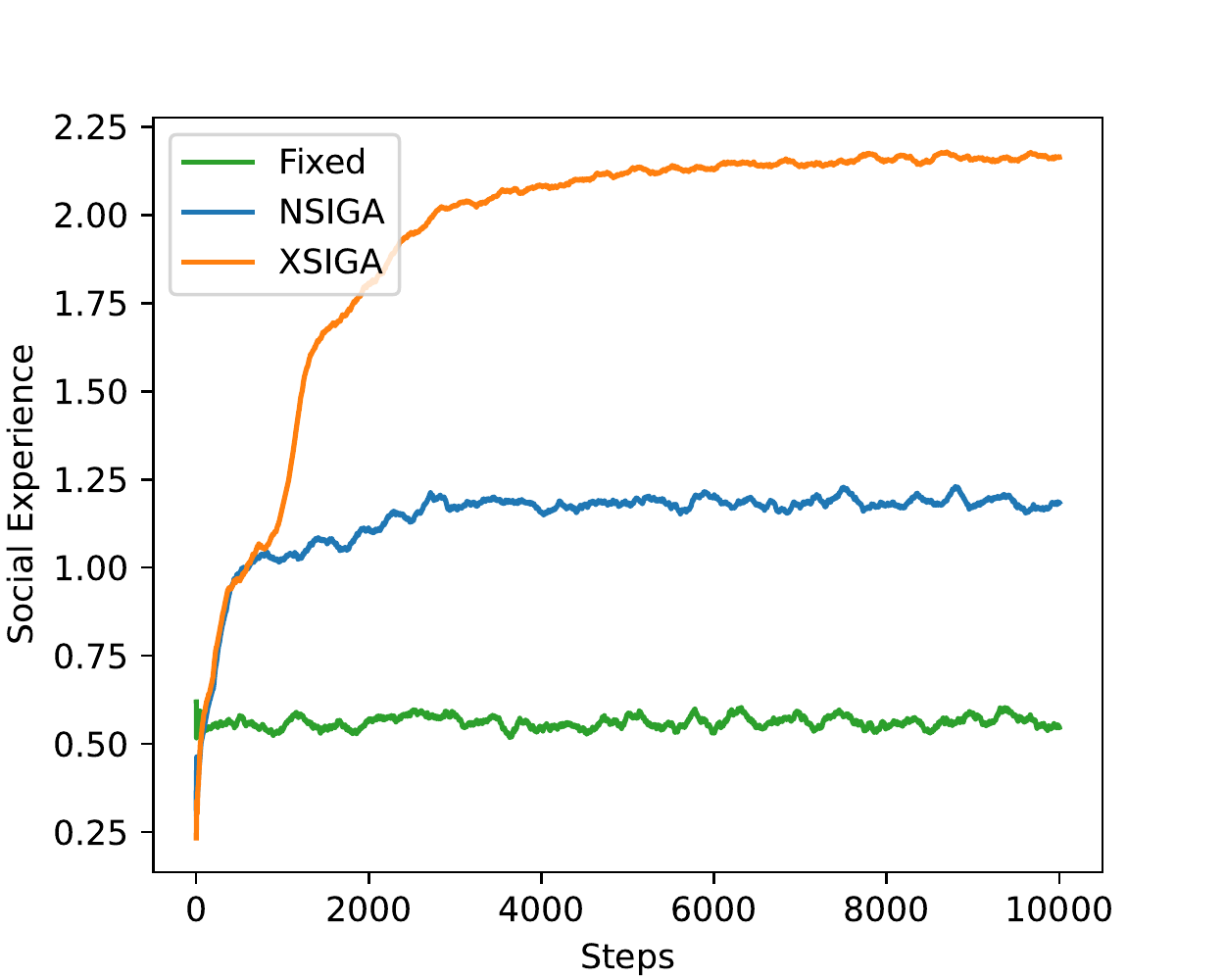}
    \caption[Social Experience in Selfish Society.]{Social Experience in Selfish Society.}
    \label{fig:social_experience_0_1_0__1_15_17_sh1}
\end{subfigure}

\begin{subfigure}[b]{0.45\textwidth}
    \includegraphics[width=\columnwidth]{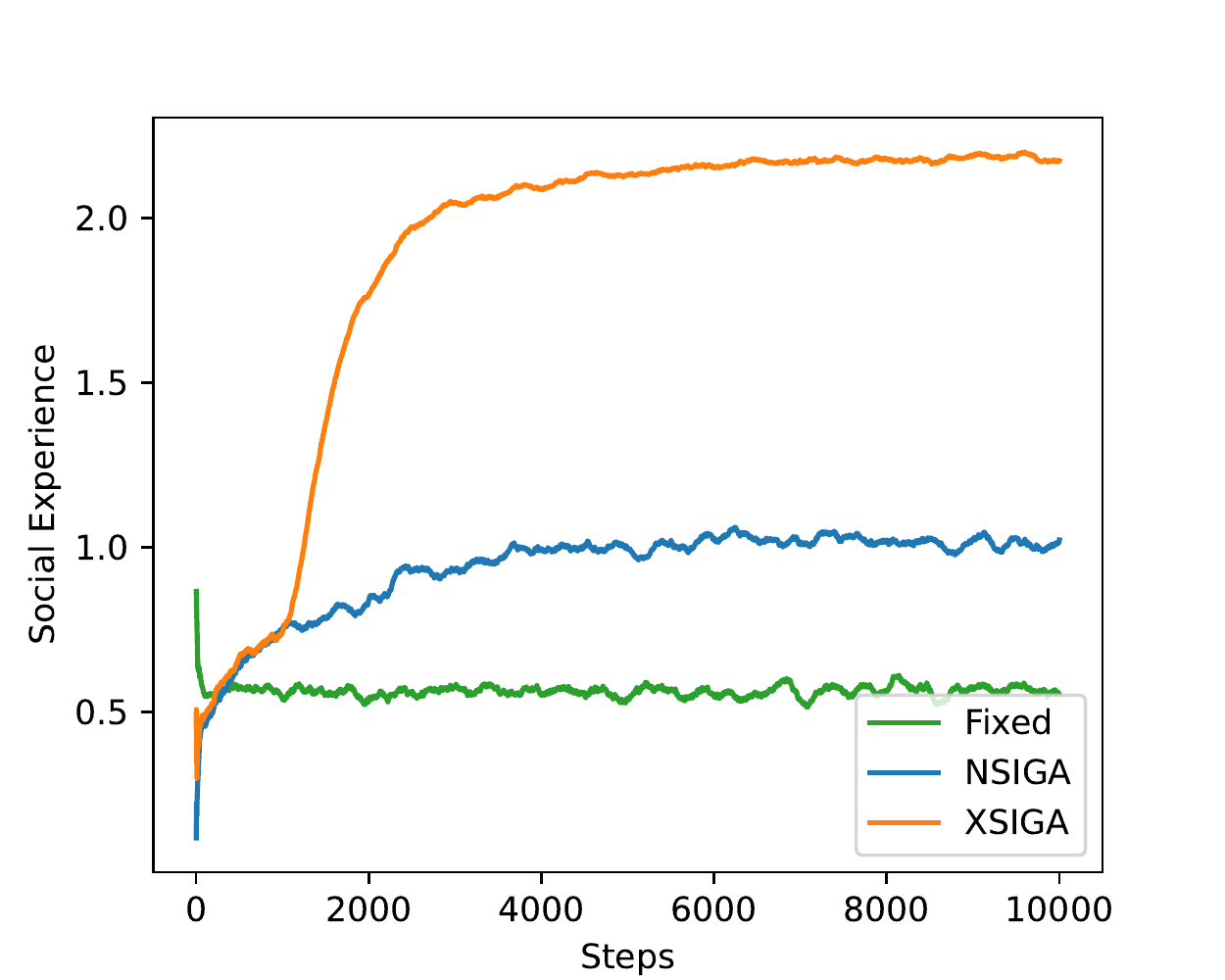}
    \caption{Social Experience in Considerate Society.}
    \label{fig:social_experience_0_0_1__1_15_17_sh1}
\end{subfigure}
\begin{subfigure}[b]{0.45\textwidth}
    \includegraphics[width=\columnwidth]{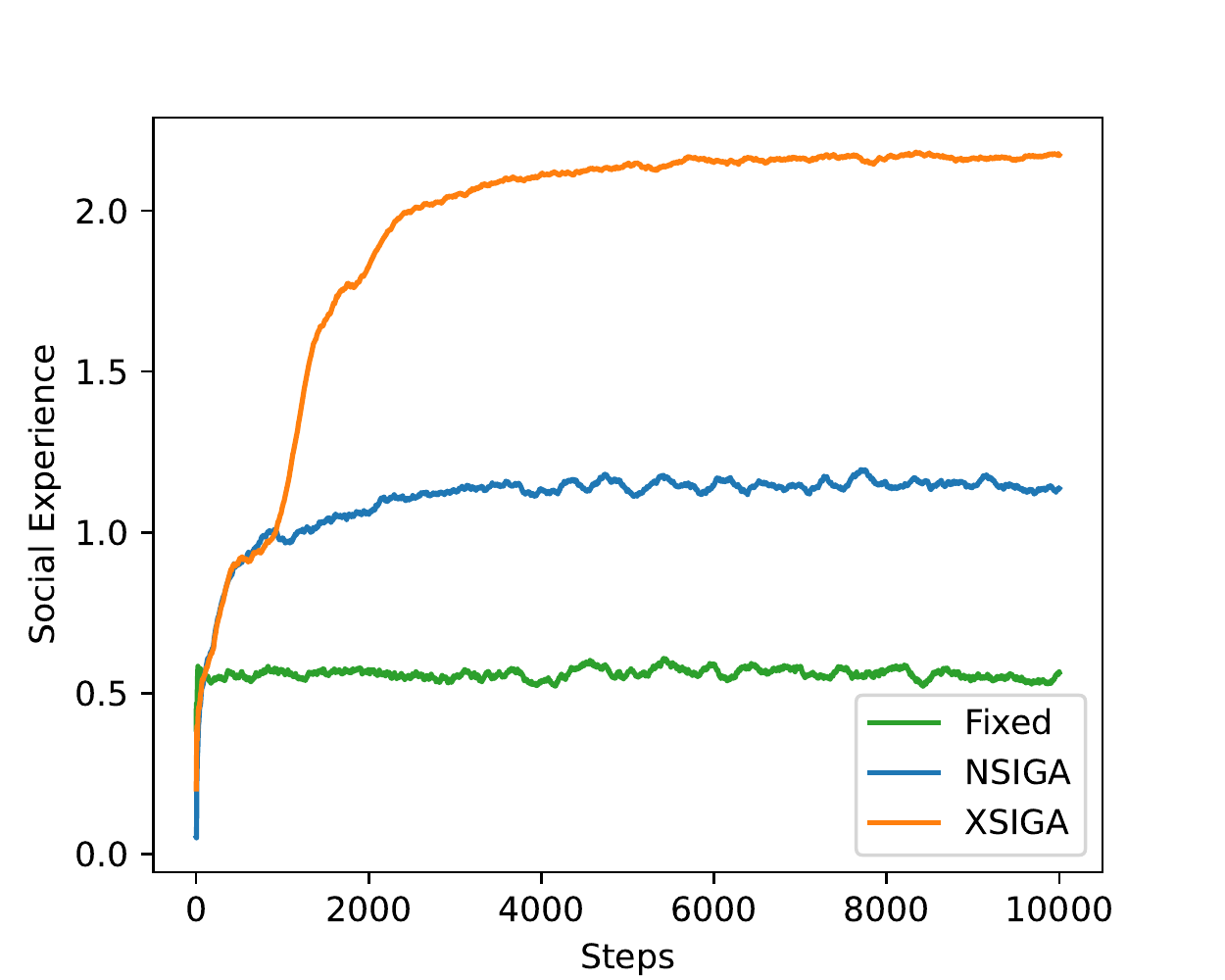}
    \caption[Social Experience in Mixed Society.]{Social Experience in Mixed Society.}
    \label{fig:social_experience_2_1_1__1_15_17_sh1}
\end{subfigure}

\caption{Social experience with payoffs in \protect\citet{Ajmeri-IJCAI18-Poros}.}
    \label{fig:social_experience-poros-payoff}
\end{figure*}

Figure~\ref{fig:social_experience-poros-payoff} shows the results for social experience in pragmatic, selfish, considerate, and mixed societies with payoffs in \citet{Ajmeri-IJCAI18-Poros}. We observe that our results hold for these payoffs.

\end{document}